\documentclass[pdflatex,sn-mathphys-num]{sn-jnl}

\usepackage{amsmath}
\usepackage{amssymb}
\usepackage{graphicx}% Include figure files
\usepackage{dcolumn}% Align table columns on decimal point
\usepackage{bm}% bold math
\usepackage{float}
\usepackage{booktabs}
\usepackage{enumerate}% http://ctan.org/pkg/enumerate
\usepackage{mathtools}
\usepackage{xcolor}

\begin{document}

\title{Revisiting the properties of superfluid and normal liquid $^4$He using {\em ab initio} potentials}

\author[1]{Tommaso Morresi}
\author*[1]{Giovanni Garberoglio}\email{garberoglio@ectstar.eu}
\affil[1]{European Centre for Theoretical Studies in Nuclear Physics and Related Areas (ECT*),
  Fondazione Bruno Kessler, Trento, Italy}

\abstract{
  We investigate the properties of liquid ${}^4$He in both the normal and superfluid phases using
  path integral Monte Carlo simulations and recently developed {\em ab initio} potentials that
  incorporate pair, three-body, and four-body interactions. By focusing on the energy per particle
  as a representative observable, we use a perturbative approach to quantify the individual
  contributions of the many-body potentials and systematically propagate their associated
  uncertainties. Our findings indicate that the three-body and four-body potentials contribute to
  the total energy by approximately 4\% and 0.4\%, respectively. However, the primary limitation in
  achieving highly accurate first-principles calculations arises from the uncertainty in the
  four-body potential, which currently dominates the propagated uncertainty. 
  In addition to the energy per particle, we analyze other key observables, including the superfluid
  fraction, condensed fraction, and pair distribution function, all of which demonstrate excellent
  agreement with experimental measurements.
}


\maketitle

\section{Introduction}

He-II, the superfluid phase of liquid ${}^4$He below $T=2.1768$~K, is the prototypical example of a
strongly-interacting quantum system whose unique properties are governed by Bose--Einstein
condensation (BEC).~\cite{Pines_Nozieres1999,Leggett2006, pitaevskii_2016}
Several theoretical approaches enabled a thorough understanding of the origin of the superfluid
properties of He-II,~\cite{Landau1947, Feynman1957} and, eventually, their relation to the
Bose--Einstein statistics. However, due to the strongly-interacting nature of this system, a
microscopic description linking the interaction between helium atoms to the thermophysical
properties of He-II can only be carried out using computer simulations. In this respect, the
path-integral formulation to quantum statistical mechanics~\cite{feynman_1965} has been proven to be
an efficient and reliable approach to compute from first principles the properties of liquid
He-II. Early calculations, using a semi-empirical form for the pair interaction between helium
atoms~\cite{kalos_1970,whitlock_1979,aziz_1979, aziz_1987}, were able to quantitatively reproduce
several properties of superfluid helium, such as the transition temperature, the specific heat, the
condensate and the superfluid fractions, with good accuracy.~\cite{ceperley_1995} 

Recently, developments in theoretical and computational quantum chemistry enabled the calculation of the pair,~\cite{czachorowski_2020} three-body~\cite{lang_2023} and four-body~\cite{wheatley_2023} potentials between helium atoms with unprecedented accuracy and well-defined uncertainty estimates. These efforts have been motivated by the realization, around the turn of the last century, that several thermophysical properties of the helium gas (notably, the coefficients of the virial expansion) could be computed {\em ab initio} with rigorously propagated uncertainties.~\cite{Aziz1995}
These uncertainties are smaller than those from the best experimental determinations. Currently, the second virial coefficient of helium is known theoretically with an uncertainty approximately 10 times smaller than the best experimental measurement.~\cite{czachorowski_2020} Similarly, the uncertainty of the third virial coefficient is about 50 times smaller than that obtained from experiments.~\cite{C_2024}

These achievements, which have been made possible by close collaboration between several experimental and theoretical groups worldwide, have had a significant impact in the development of primary standards of temperature using gas thermometry. The state of the art regarding the use of first-principles calculations in temperature and pressure gas-based metrology has been summarized in a recent review.~\cite{garberoglio_2023_review}

Despite the successes of using {\em ab initio} potentials to describing the properties of helium in the gas phase, comparably less efforts have been devoted to investigate the accuracy with which thermophysical properties of condensed phase can be described from first principles.
Almost all of the calculations found in the literature, even those appeared in recent years (see for example, Refs.~\cite{Boninsegni_23}), still use some of the pair potentials developed by Aziz (mainly the original 1979 version~\cite{aziz_1979}, the 1987 revision~\cite{aziz_1987} -- also known as HFD-B(He) potential -- or the {\em ab initio} 1995 implementation~\cite{Aziz1995}) and neglect three-body contributions, generally finding good agreement with experimental data.

Recently, mainly driven by the interest in the phenomenon of supersolidity, some studies assessed the effect of three-body forces, both using a semiclassical approach~\cite{Xu24} or path-integral simulations in the solid phase.~\cite{Barnes_17}
Additionally, {\em ab initio} pair and three-body potentials have been fully taken into account also in simulations of supercritical helium above the vapor--liquid critical point.~\cite{marienhagen_2024}
These studies generally found that three-body forces are needed if accuracy of the order of few percent or less is required and point out that accurate thermophysical properties of condensed helium can indeed be obtained by a fully {\em ab initio} approach. However, it is not yet clear whether the same outstanding performance of {\em ab initio} potentials that has been exploited in primary temperature metrology~\cite{garberoglio_2023_review} can be transferred to the liquid phase. It is worth noting that the most used temperature scale (ITS-90) is based on the vapor pressure of helium when covering the range $0.65-5$~K;~\cite{ITS90} given the high accuracy achieved in determining the virial coefficients, it is natural to wonder how well this quantity could be computed from first principles.

In this work, we set the stage for investigating if and how one can compute thermophysical properties of liquid helium (normal and superfluid) from first principles with well-defined uncertainties.
We aim at determining whether the precision achieved in {\em ab initio} calculation of helium properties in the gas phase can be extended to the liquid, with a focus on quantifying the impact of higher-order many-body interactions on the energy of the system.
To this end, we critically examine all the (inevitable) choices that one has to make while performing computer simulations and estimate their effect on the quality of the results. 
We implemented the most recent Path-Integral Monte Carlo (PIMC) algorithm that takes into account exchange effects between ${}^4$He atoms in the condensed phase,~\cite{prokofev_1998,proko_2001,boninsegni_2006,Spada_2022} and used it to simulate liquid helium across the superfluid transition using the most recent pair potential available in the literature.~\cite{czachorowski_2020} We computed the kinetic and potential energy as a function of temperature, as well as the pair distribution function, the condensate fraction and the superfluid fraction, finding a very good agreement with experimental data.

Although path-integral simulations of liquid ${}^4$He can be performed including higher-order many-body potentials, this kind of calculation would presently require a considerable computational effort. 
Therefore, we decided to use a perturbative approach to evaluate the effect of three-body~\cite{lang_2023} and four-body interactions,~\cite{wheatley_2023} starting from configurations generated using only the pair-potential. In this case, one can use an efficient simulation algorithm -- known as the {\em pair product approximation} -- to sample the relevant configurations.
This approach of perturbative evaluation of the many-body effect in condensed quantum systems has been shown to provide very good estimates when used to investigate the properties of hydrogen,~\cite{Ibrahim2022} and we deemed it appropriate for our aims. Our results can be used as a stepping stone for more accurate, and computationally demanding, simulations aimed at computing thermophysical properties with uncertainties determined by the propagated uncertainties from the potentials and not the statistical uncertainty of the Monte Carlo calculation. Additionally, this work will facilitate the estimation of computational requirements for more complex simulations, allowing for better planning and resource allocation.

We found that the effect of three-body interactions on the energy per particle of liquid ${}^4$He is $\sim 4\%$, while the four-body potential results in an additional change of $\sim 0.4\%$. We also propagated the uncertainty assigned to the many-body potentials to the expected uncertainty of the energy. The effect of the pair potential is negligible, while the three-body potential is on the order of one part in $10^4$. The uncertainty propagated from the four-body potential is roughly 50\% of the energy contribution due to the four-body potential ({\em i.e.}, 0.2\% of the total energy), and this is, at the moment, the limiting factor in determining the theoretical accuracy of the energy per particle in the liquid phase.
% ERRORi in K: (2-BODY) (3-BODY) (4-BODY)
% 0.0000 & 0.0015 & 0.020 \\

Our results compare very well with experimental values in the superfluid regime, while we observe a systematic discrepancy in the normal liquid phase; we argue that this difference might be due to limitations of the pair-product approximation. We will highlight critical areas for methodological improvement, paving the way for more precise simulations of quantum fluids and optimization of computational resources to achieve uncertainty levels commensurate with experimental precision.

The paper is organised as follows: in section \ref{sec:methods} we describe the general theoretical
framework that we used to perform the simulations; in section \ref{sec:details} we give the
computational details of our calculations, and in section \ref{sec:estimators} we discuss how the
observable quantities have been estimated. Finally, in section \ref{sec:results}, we discuss our
results before presenting some concluding remarks in section \ref{sec:concl}.

\section{Methods}\label{sec:methods}
We consider a system of $N$ $^4$He particles described by the Hamiltonian $H=K+V$, where $K$ is the kinetic part and $V$ is the potential one, and with inverse temperature $\beta=\frac{1}{k_\mathrm{B} T}$, where $k_\mathrm{B}$ is the Boltzmann constant. The canonical partition function $Z$ of this system, in the coordinate representation, is given by
\begin{equation}
  Z=\frac{1}{N!}\sum_{\mathcal{P}} \int d \mathbf{R} \ \rho \left( \mathbf{R},\mathcal{P} \mathbf{R},\beta \right),
\end{equation}
where $\mathbf{R}=\left(\mathbf{r}_1,\mathbf{r}_2,\dots,\mathbf{r}_N \right)$ is a compact notation for all the particle coordinates, $\rho(\mathbf{R},\mathbf{R}',\beta)=\langle \mathbf{R}|\text{e}^{-\beta H} | \mathbf{R}' \rangle$ is the canonical density matrix and $\mathcal{P} \mathbf{R}=(\mathbf{r}_{\mathcal{P}(1)},\mathbf{r}_{\mathcal{P}(2)},\dots,\mathbf{r}_{\mathcal{P}(N)})$ denotes the position vectors with permuted labels. 
Considering a cubic box of volume $\Omega=L^3$ with periodic boundary conditions, one can generalise the notation by calling $\mathbf{X}=(\mathbf{x}_1,\mathbf{x}_2,\dots,\mathbf{x}_N)$ the coordinates inside the unit cell, and with $\mathbf{R}=(\mathbf{x}_1+\mathbf{w}_1L,\mathbf{x}_2+\mathbf{w}_2L,\dots,\mathbf{x}_N+\mathbf{w}_NL)$ the general coordinates, with $\mathbf{W}=(\mathbf{w}_1,\mathbf{w}_2,\dots,\mathbf{w}_N)$ a vector of integers which can take any value in ($-\infty$,$\infty$). Therefore we will use $| \mathbf{R} \rangle = | \mathbf{X}, \mathbf{W} \rangle$ from now on.
In the case that the potential $V$ is not long-ranged, {\em i.e.}, it falls to zero inside the simulation box $\Omega$, the partition function of such a system can be rewritten as
\begin{equation}\label{eq:z}
  Z=\frac{1}{N!}\sum_{\mathcal{P}} \sum_{\mathbf{W}} \int_{\Omega} \text{d} \mathbf{X} \ \langle \mathbf{X}, 0 | \text{e}^{- \beta H} | \mathcal{P} \mathbf{X}, \mathbf{W} \rangle.
\end{equation}
By using the convolution property and considering that we are dealing with a bosonic system, one can further write
\begin{equation}\label{eq:zconv}
  \begin{split}
    Z &= \frac{1}{N!} \sum_{\mathcal{P}} \sum_{\mathbf{W}} \int_{\Omega} \text{d} \mathbf{X}^{(0)} \int \text{d}
    \mathbf{R}^{(1)} \dots \text{d} \mathbf{R}^{(P-1)}   
    \rho \left(\mathbf{X}^{(0)},\mathbf{R}^{(1)},\tau \right) \rho
    \left(\mathbf{R}^{(1)},\mathbf{R}^{(2)},\tau \right) \dots \\
    & \rho(\mathbf{R}^{(P-1)},\mathcal{P}\mathbf{X}^{(0)}+\mathbf{W}L,\tau),
  \end{split}
\end{equation}
where $\tau=\frac{\beta}{P}$ and $P$ is the number of replica (beads).
It is worth to emphasize that while the vectors $\mathbf{R}^{(1)},\dots,\mathbf{R}^{(P)}$ span the entire space, the vector $\mathbf{X}^{(0)}$ spans the unit cell only. 

The main idea of PIMC is to exploit Eq. (\ref{eq:zconv}) to perform a classical-like simulation of $N$ polymers formed by $P$ links, connecting $P+1$ beads. Each polymer can either close on itself (if $\mathcal{P}(i)=i$) or on another particle specified by the permutation index $\mathcal{P}(i)$. 
In this work, we use the PIMC worm algorithm as implemented in Ref.~\cite{Spada_2022} which is rigorously compatible with the periodic boundary conditions used in liquid simulations and fixes some ambiguities of the original version that arise when the de Broglie thermal wavelength becomes comparable to the size of the simulation box. This algorithm enables sampling the effect of the permutation operators $\cal P$, which play a major role in the superfluid regime of $^4$He.
Within this approach, the bead $0$ of the polymer corresponding to particle $i$ is confined within the simulation box, while the endpoint at bead $P$ connects to an image of the first bead of the polymer $\mathcal{P}(i)$, identified up to the periodicity of the simulation box. Therefore, in Eq.~(\ref{eq:zconv}), the first bead belongs to the simulation box, while the others are free to move without restrictions.
Nevertheless, particle interactions are always computed using the minimum image convention.

The key quantity in a PIMC simulation is the propagator $\rho(\mathbf{R}^{(j)},\mathbf{R}^{(j+1)},\tau)$. In order to evaluate this quantity, different approximations are available.~\cite{ceperley_1995} 
The most common, simplest, and controlled approximation is the primitive factorisation of the propagator, such that
\begin{equation}\label{eq:primitive_approx}
  \langle \mathbf{X}, 0 | \text{e}^{- \tau H} | \mathcal{P} \mathbf{X}, \mathbf{W} \rangle \eqsim \langle \mathbf{X}, 0 | \text{e}^{- \tau K} \text{e}^{- \tau V}  | \mathcal{P} \mathbf{X}, \mathbf{W} \rangle.
\end{equation}
It is based on the identity $ \text{e}^{-\tau(K+V)+\frac{\tau^2}{2}\left[ K,V\right]}= \text{e}^{-\tau K} \text{e}^{-\tau V}$.
Indeed, if P is large enough the term with the commutator $[K,V]$ goes to zero quadratically and can be neglected. This approximation allows the evaluation of the exact density matrix in terms of two separate density matrices, the kinetic one and the potential one. The latter is diagonal in the position representation and it is trivially evaluated as
\begin{equation}\label{eq:vprim}
  \left\langle \mathbf{R}^{(j)} \left| \text{e}^{-\tau V} \right| \mathbf{R}^{(j+1)} \right\rangle = \text{e}^{-\tau V(\mathbf{R}^{(j)})} \delta \left( \mathbf{R}^{(j)} - \mathbf{R}^{(j+1)} \right).
\end{equation}
The kinetic part is instead computed as~\cite{ceperley_1995}
\begin{equation}
  \left\langle \mathbf{R}^{(j)} \left| \text{e}^{-\tau T} \right| \mathbf{R}^{(j+1)} \right\rangle = \left(4 \pi \lambda \tau \right)^{\frac{3N}{2}} \text{e}^{-\frac{\left( \mathbf{R}^{(j)} -\mathbf{R}^{(j+1)}\right)^2}{4\tau \lambda}}, 
\end{equation}
where $\lambda=\frac{\hbar^2}{2m}$ and $m$ is the mass of the $^4$He atom.
Using a general notation, the density matrix can be written as
\begin{equation}\label{eq:rho_general}
  \rho \left(\mathbf{R}^{(j)},\mathbf{R}^{(j+1)},\tau \right) = \rho_0 \left(\mathbf{R}^{(j)},\mathbf{R}^{(j+1)},\tau \right) \ \exp \left[{-U \left(\mathbf{R}^{(j)},\mathbf{R}^{(j+1)},\tau \right)} \right],
\end{equation}
where 
\begin{equation}
  \rho_0\left(\mathbf{R}^{(j)},\mathbf{R}^{(j+1)},\tau \right) = \prod_{i=1}^N \rho^\mathrm{sp}_{0} (\mathbf{r}^{(j)}_{i},\mathbf{r}^{(j+1)}_{i},\tau) = \frac{1}{\left( 4 \pi \lambda \tau  \right)^{\frac{3N}{2}}} \prod_{i=1}^N \exp \left[ - \frac{(\mathbf{r}^{(j)}_{i}-\mathbf{r}^{(j+1)}_{i})^2}{4 \lambda \tau} \right], 
\end{equation}
and in the primitive case the interacting part of the action is evaluated as
\begin{equation}
  U\left(\mathbf{R}^{(j)},
  \mathbf{R}^{(j+1)},\tau \right) = \frac{\tau \left( V \left(\mathbf{R}^{(j)} \right) + V \left(\mathbf{R}^{(j+1)} \right) \right)}{2}.
\end{equation}

If $V$ is approximated as a sum of two-body potential terms ($V \left( \mathbf{r}_1,...,\mathbf{r}_N\right)\equiv \sum_{i_1,i_2} V_2 (\mathbf{r}_{i_1}-\mathbf{r}_{i_2})$), a more efficient framework to estimate the propagator is the pair-product approximation;~\cite{Pollock_1984} this approach is very efficient, at the expense of some uncontrolled approximations. The main idea behind this approach is that one is able to solve the two-body problem (numerically) exactly. Therefore one can build the exact propagator for two atoms, and then use the latter to approximate the many-body propagator. In particular, using the Feynman--Kac formula,~\cite{ceperley_1995} the interacting part of the propagator in Eq.~(\ref{eq:rho_general}) can be exactly computed as
\begin{equation}\label{eq:FK}
  \text{e}^{ {-U \left(\mathbf{R}^{(j)},\mathbf{R}^{(j+1)},\tau \right)}} \equiv \left\langle \prod_{i_1<i_2} \text{e}^{-\int_0^{\tau} V_2(\mathbf{r}_{i_1 i_2}(\tau')) d\tau'} \right\rangle_{RW},
\end{equation}
where the brackets stand for an average over all the random walks (RW) of the free particles from $\tau=0$ to $\tau'=\tau$ and $\mathbf{r}_{i_1i_2}$ is the relative distance between particles labelled by $i_1$ and $i_2$. If the terms of the product on the right hand side of 
Eq.~(\ref{eq:FK}) are uncorrelated, one can interchange the product and the averaging operation such that
\begin{equation}\label{eq:FK_approx}
  \text{e}^{ {-U \left(\mathbf{R}^{(j)},\mathbf{R}^{(j+1)},\tau \right)}} \eqsim  \prod_{i_1<i_2} \left\langle \text{e}^{-\int_0^{\tau} V_2(\mathbf{r}_{i_1i_2}(\tau')) d\tau'} \right\rangle_{RW} \equiv \prod_{i_1<i_2} \left\langle \text{e}^{- U_2\left(\mathbf{r}_{i_1i_2},\mathbf{r}'_{i_1i_2},\tau \right)} \right\rangle,
\end{equation}
where $U_2\left(\mathbf{r}_{i_1i_2},\mathbf{r}'_{i_1i_2},\tau \right)$ can be computed (numerically) exactly for atoms $i_1$ and $i_2$. Although the resulting propagator needs a much smaller number of beads to achieve convergence, this method relies the uncontrolled approximation of Eq.~(\ref{eq:FK_approx}). As we show in the Appendix, it is difficult to estimate the magnitude of this additional uncertainty in the energy per particle, but we argue that it should not exceed $0.1$~K. %sacrificing accuracy.  
Within this approach, the scheme to rewrite $\rho(\mathbf{R}^{(j)},\mathbf{R}^{(j+1)},\tau)$ in Eq.~(\ref{eq:zconv}) is \cite{Pollock_1984,ceperley_1995}
\begin{equation}\label{eq:rho_pp}
  \rho(\mathbf{R}^{(j)},\mathbf{R}^{(j+1)},\tau) = \left[ \prod_{i=1}^N \rho^\mathrm{sp}_{0} (\mathbf{r}^{(j)}_{i},\mathbf{r}^{(j+1)}_{i},\tau) \right] \cdot  \left[ \prod_{i_1 < i_2}^N \frac{ \rho^\mathrm{rel}\left( \mathbf{r}^{(j)}_{i_1 i_2},\mathbf{r}^{(j+1)}_{i_1i_2},\tau \right)}{\rho_0^\mathrm{rel}\left( \mathbf{r}^{(j)}_{i_1i_2},\mathbf{r}^{(j+1)}_{i_1i_2},\tau \right)} \right],  
\end{equation}
where $\rho^\mathrm{sp}_0$ is the single-particle ideal-gas density matrix, $\rho^\mathrm{rel}_{0}(\mathbf{r}^{(j)}_{i_1i_2},\mathbf{r}^{(j+1)}_{i_1i_2},\tau) = \frac{1}{\left( 8 \pi \lambda \tau  \right)^{\frac{3}{2}}} \exp \left[ - \frac{(\mathbf{r}^{(j)}_{i_1i_2}-\mathbf{r}^{(j+1)}_{i_1i_2})^2}{8 \lambda \tau} \right]$, and where
\begin{align}\label{eq:rho_num}
  \rho^\mathrm{rel}\left( \mathbf{r}^{(j)}_{i_1i_2},\mathbf{r}^{(j+1)}_{i_1i_2},\tau \right) &=
  \sum_{l=0}^{\infty} \frac{2l+1}{4 \pi} P_l(\cos(\theta)) \left[ \sum_b \langle
    r^{(j)}_{i_1i_2}|\psi_{b,l} \rangle \langle \psi_{b,l} | r^{(j+1)}_{i_1i_2} \rangle
    \text{e}^{-\tau E_{b,l}} + \right. \nonumber \\
    & \left. \int \text{d}k \langle r^{(j)}_{i_1i_2}|\psi_{k,l} \rangle \langle \psi_{k,l} |
    r^{(j+1)}_{i_1i_2} \rangle \text{e}^{-\tau E_{k,l}} \right], 
\end{align}
where $\cos(\theta)$ is the cosine of the angle between vectors $\mathbf{r}^{(j)}_{i_1i_2}$ and $\mathbf{r}^{(j+1)}_{i_1i_2}$ in two consecutive imaginary-time slices ($j$-th and $j+1$-th), the sum inside parentheses runs over bound states and the integral is over continuum states.
The $\psi_{b/c,l}$ denotes the bound and continuum solutions of the radial Schr{\"o}dinger equation with angular momentum $l$ for the given two-body potential $V_2$.  
Finally, the resulting interacting part of the action to be inserted in Eq.~(\ref{eq:rho_general}) is thus 
\begin{equation}
  U\left(\mathbf{R}^{(j)},\mathbf{R}^{(j+1)},\tau \right) = - \sum_{i_1<i_2} \log\left[ \frac{ \rho^\mathrm{rel}\left( \mathbf{r}^{(j)}_{i_1 i_2},\mathbf{r}^{(j+1)}_{i_1i_2},\tau \right)}{\rho_0^\mathrm{rel}\left( \mathbf{r}^{(j)}_{i_1i_2},\mathbf{r}^{(j+1)}_{i_1i_2},\tau \right)} \right].
\end{equation}

All the results of the present work have been obtained using the pair-product approximation. In section \ref{sec:pp_vs_primitive} of the Appendix we show the advantage of using the pair-product approximation with respect to the primitive one by analysing the convergence of the energy-per-particle with respect to the number of replicas.

\subsection{PIMC moves}
The extended configuration space containing $NP$ particles is sampled through a Monte Carlo integration scheme based on the worm algorithm \cite{prokofev_1998,boninsegni_2006} and the Metropolis--Hastings acceptance/rejection.~\cite{metropolis_1953,hastings_1970} 
In this framework, observables are sampled according to Eq.~(\ref{eq:zconv}), which is called the Z-configuration (diagonal) space. The permutations $\cal P$ are sampled in the so-called G-configuration (off-diagonal) space, obtained by cutting one of the cycles and leaving one sequence open-ended. 
The algorithm allows correct sampling of both the diagonal and off-diagonal sectors.~\cite{Spada_2022} Here we quickly remind the original set of moves, that are:
\begin{itemize}
\item \textit{Translate}: this move generates a rigid displacement for a single polymer;   
\item \textit{Redraw}: this move selects a part of a polymer, {\em i.e.}, some of its beads, and resamples it by the Levy construction as reported in Ref. \cite{fosdick_1966};
\item \textit{Open / Close}: these two moves allow to switch from the diagonal sector Z to the off-diagonal one G and vice versa. The open move cuts a link of a polymer in the P-th time slice, creating the worm which is characterized by a head and a tail, while the close move binds the two ends of the worm;
\item \textit{Swap}: this is the move enabling the sampling of the permutation space by exchanging the worm head with another polymer. It can be applied only when the system is in the G-space;
\item \textit{Move head}: this move can be applied in the G-space and it samples a part of the worm using the Levy reconstruction;
\item \textit{Move tail}: this move can be applied in the G-space and it samples a part of the tail using again the Levy reconstruction.
\end{itemize}
In this work, we introduce a new global move in Z-space, referred to as {\em Shake}, building upon the set proposed in Ref.~\cite{Spada_2022}. This move leverages normal mode sampling to efficiently decorrelate the generated Monte Carlo configurations.
The main idea behind it comes by noticing that, for any possible permutation $\cal P$, the kinetic-energy term in the primitive approximation corresponds to a set of harmonic oscillators, that is~\cite{ceriotti_2010}
\begin{align}\label{eq:HPgen}
  K_P(\mathbf{R}, \mathbf{P}) &= \sum_{n=1}^{N_{pol}} \left\{ \sum_{i=1}^{N_n} \left[
    \sum_{j=0}^{P-2} \left( \frac{\left[\mathbf{p}^{(j)}_{s(i)}\right]^2}{2m} +
    \frac{1}{2}m\omega_P^2\left(\mathbf{r}^{(j)}_{s(i)} -\mathbf{r}^{(j+1)}_{s(i)} \right)^2
    \right) + \frac{\left[\mathbf{p}^{(P-1)}_{s(i)}\right]^2}{2m_{s(i)}} + \right. \right. \nonumber \\
    & \left. \left. \frac{1}{2}m_{s(i)}\omega_P^2 \left(\mathbf{r}^{(P-1)}_{s(i)}
    -\mathbf{r}^{(0)}_{\mathcal{P}(s(i))}\right)^2 \right] \right\},  
\end{align}
where $s$ is the vector of length $N_n$ of atomic indices labeling the atoms of the $n$-th polymer, $N_n$ is the number of atoms in the $n$-th polymer, so that $\sum_{n=1}^{N_{pol}} N_n \equiv N$, $\omega_P = \frac{1}{\tau \hbar}$ and $N_{pol}$ is the number of polymers in the current configuration. 
In particular, one can exploit the knowledge of the normal modes of $K_P$ which are known analytically~\cite{ceriotti_2010} even in the case that polymers are connected: 
\begin{equation}
  C^{(n)}_{jk}=  
  \begin{dcases}
    \left[\sqrt{\frac{2}{N_nP}}-\left( \frac{\sqrt{2}-1}{\sqrt{N_nP}}\right)\left( \delta_{k,0}+\delta_{k,\text{int}\left[\frac{N_nP+1}{2}\right]}\right)\right] \  \text{cos}\left( \frac{2 \pi j k}{N_nP}\right),              & \text{if } 0 \le k \le \text{int}\left[\frac{N_nP}{2}\right],\\
    \sqrt{\frac{2}{N_nP}} \  \text{sin}\left( \frac{2 \pi j k}{N_nP}\right),              & \text{if } \text{int}\left[\frac{N_nP}{2}\right]+ 1 \le k \le N_nP-1,
  \end{dcases}  
  \label{eq:C}
\end{equation}
where $\text{int}\left[x\right]$ gives the largest integer smaller than or equal to $x$.
In Eq.~(\ref{eq:C}), the quantity $C^{(n)}_{jk}$ is proportional to the amplitude of the motion of bead $k$ in the $j$-th mode of the $n$-th polymer in the system.
Similarly to what is done in Refs.~\cite{takahashi_1984,runge_1988}, we use this expression to update the configuration of each polymer of the system according to
\begin{equation}
  \left( \bar{\mathbf{r}}_{s(1)},\bar{\mathbf{r}}_{s(2)},\dots,\bar{\mathbf{r}}_{s(N_n)} \right) \to \left( \bar{\mathbf{r}}_{s(1)},\bar{\mathbf{r}}_{s(2)},\dots,\bar{\mathbf{r}}_{s(N_n)} \right) + \sum_{a=1}^3\sum_{k=0}^{N_nP-1} C^{(n)}_{j_{R}k} \cdot \Delta^{(n)}_{ka},
\end{equation}
where $\bar{\mathbf{r}}_{s(i)}$ is a compact notation for the $3P$-dimensional vector representing the positions of all the beads of the $s(i)$-th atom,  %in the $n$-th polymer, 
$j_R$ is a random mode which is chosen with equal probability among the $N_nP$ modes available for polymer $n$ and $\Delta^{(n)}_{ka}$ is a $N_nP$-dimensional vector of random displacements along the $a$-axis. The move is then accepted or rejected with probability $A_S$:
\begin{equation}
  A_S = \text{min}\left\{ 1,\text{exp} \left[ \sum_{i=1}^{N}\sum_{j=1}^{P} \left( U\left(\mathbf{R}^{(j)},\mathbf{R}^{(j+1)},\tau \right) - U\left(\mathbf{R}'^{(j)},\mathbf{R}'^{(j+1)},\tau \right) \right)  \right] \right\},
\end{equation}
and the magnitude of $\Delta^{(n)}_{ka}$ is chosen during the equilibration phase in order to get an acceptance ration close to $\frac{1}{2}$. 

\section{Computational details}\label{sec:details}
In all our simulations, the density was fixed to reproduce, for any given temperature, the experimental value along the vapor-liquid coexistence curve. In particular, we used the values from Ref.~\cite{brooks_1977} for temperatures below 2.1~K and those from Ref.~\cite{crawford_1977} for temperatures between 2.1~K and 4~K.

We used the pair-product approximation for the propagator with a fixed value of $\tau=\frac{1}{40}$
[K$^{-1}$] (see the Appendix for the convergence analysis). The number of beads for each temperature
$T$ was thus chosen as $P=\mathrm{int} \left[\frac{1}{\tau T} \right]$. In our calculations, only
the pair potential was used to sample the liquid configurations. The three-body and four-body
potentials could, in principle, be used in the sampling procedure, but at the cost of a considerable
increase in computational requirements. A straightforward way to do so would be to resort to the
primitive approximation; however, this approach would immediately result in an order-of-magnitude
increase in computational requirements. We performed some test calculations within this framework,
and our results indicate that using the three-body potential will further require a 30-fold increase
in computational resources. Furthermore, in its present form, the evaluation of the four-body
potential energy surface is quite slow, and its explicit inclusion would increase the computational
cost by an additional factor of $10^4$. These huge increments justify our choice of using the
pair-product approximation for sampling configurations and the subsequent perturbative evaluation of
the three- and four-body terms in this exploratory work. We discuss below strategies for more
accurate future calculations. 

In Eq.~(\ref{eq:rho_num}) the radial continuum wavefunctions are computed by integrating the Schr{\"o}dinger equation for the given 2-body potential (Aziz or Czachorowski) using the Numerov method and $1.6 \times 10^4$ points from $r_{m} = 0.05 \ \text{\AA}$ to $r_M = 60.05 \ \text{\AA}$. Once fixed $\tau$, due to the spherical symmetry of the 2-body problem, $\rho^\mathrm{rel}\left( \mathbf{r}^{(j)}_{i_1i_2},\mathbf{r}^{(j+1)}_{i_1i_2},\tau \right)$ depends only on three variables which are $|\mathbf{r}^{(j)}_{i_1i_2}|$, $|\mathbf{r}^{(j+1)}_{i_1i_2}|$ and the cosine of the relative angle $\theta$ between the $\mathbf{r}^{(j)}_{i_1i_2}$ and $\mathbf{r}^{(j+1)}_{i_1i_2}$. The value of $\rho^\mathrm{rel}$ has been tabulated using a 3-dimensional grid of $104\times 104 \times 75$ points, where the two radial coordinates belong to the range [1.1, 5.04]~\text{\AA} \  and cos($\theta$) to the range [0.3, 1.0]. If the two vectors $\mathbf{r}^{(j)}_{i_1i_2}$ and $\mathbf{r}^{(j+1)}_{i_1i_2}$ are such that they fall outside the grid, the propagator is evaluated using the end-point approximation,~\cite{Pollock_1984} for which $\rho^\mathrm{rel}\left( \mathbf{r}^{(j)}_{i_1i_2},\mathbf{r}^{(j+1)}_{i_1i_2},\tau \right)=\exp\left[ -\frac{U\left(|\mathbf{r}^{(j)}_{i_1i_2}|,\tau\right)+U\left(|\mathbf{r}^{(j+1)}_{i_1i_2}|,\tau\right)}{2} \right]$, where $U\left(|\mathbf{r}^{(j)}_{i_1i_2}|,\tau\right)=-\log\left[ \frac{ \rho^\mathrm{rel}\left( \mathbf{r}^{(j)}_{i_1 i_2},\mathbf{r}^{(j)}_{i_1i_2},\tau \right)}{\rho_0^\mathrm{rel}\left( \mathbf{r}^{(j)}_{i_1i_2},\mathbf{r}^{(j)}_{i_1i_2},\tau \right)} \right]$.

\section{Estimators}\label{sec:estimators}
\subsection{Energy and specific heat}
The virial estimator for the energy is computed as~\cite{ceperley_1995,Spada_2022}:
\begin{equation}\label{eq:ev}
  \begin{split}
    E_V = \Big\langle \frac{ND}{2\beta} &+\frac{\left( \mathbf{R}^{(P-1)}-\mathbf{R}^{(P)}\right)\cdot \left( \mathbf{R}^{(P)}-\mathbf{R}^{(0)}\right)}{4\lambda\tau^2P} 
    + \frac{1}{P} \sum_{j=0}^{P-1} \frac{\partial U(\mathbf{R}^{(j)},\mathbf{R}^{(j+1)})}{\partial \tau} \\
    &+ \frac{1}{2\beta} \sum_{j=0}^{P-1} \left( \mathbf{R}^{(j)} -\mathbf{R}^{(0)} \right) \cdot \frac{\partial}{\partial \mathbf{R}^{(j)}} \left[ U(\mathbf{R}^{(j-1)},\mathbf{R}^{(j)})+U(\mathbf{R}^{(j)},\mathbf{R}^{(j+1)}) \right] \Big\rangle,    
  \end{split}
\end{equation}
where the brackets imply the average over the configurations generated through PIMC. 
The advantage of using the virial estimator instead of the thermodynamical one, which is obtained by
differentiating the partition function $Z$ in Eq.~(\ref{eq:zconv}) with respect to $\beta$:
$E_T=-\frac{1}{Z}\frac{dZ}{d\beta}$, is that its variance is much
lower.~\cite{ceperley_1995,janke_1997}   
The virial kinetic energy is simply $ K_V=E_V- \mathcal{V}_2 $,~\cite{filippi_1998} where $\mathcal{
  V}_2$ is the potential energy estimator $\mathcal{ V}_2 = \left\langle \frac{1}{P}\sum_{j=0}^{P-1}
V_2 \left(\mathbf{r}^{(j)}_1,\dots,\mathbf{r}^{(j)}_N \right) \right\rangle$. 

In this paper, the three-body $V_3$ and four-body $V_4$ potentials are not included in the propagator. Rather they are evaluated perturbatively on the configurations generated through the two-body propagator acceptance/rejection algorithm. Therefore,  $\mathcal{ V}_{3/4} = \left\langle \frac{1}{P}\sum_{j=0}^{P-1} V_{3/4} \left(\mathbf{r}^{(j)}_1,\dots,\mathbf{r}^{(j)}_N \right) \right\rangle$.

The constant volume heat capacity (C$_{\Omega}$) is the main signature of the superfluid phase transition in ${}^4$He at T$_{\lambda}$. The characteristic shape of the experimental curve is the origin of the name "$\lambda$-transition". While explicit estimators for C$_{\Omega}$ are known,~\cite{glaesemann_2002,shiga_2005} their variance grows like $N$ in the virial case and it is very hard to converge below T$_{\lambda}$. Double virial heat capacity estimators eliminate this growth of the variance but they require the calculation of the second derivative of the potential. Therefore, here we use the simple numerical differentiation of the energy: $C_{\Omega}=-\frac{\partial E_V}{\partial T}$, where $E_V$ is computed as in Eq.~(\ref{eq:ev}).

\subsection{Superfluid density}
For the calculation of the superfluid density we use the estimator proposed in Ref.~\cite{pollock_1987}, which reads as
\begin{equation}\label{eq:sd}
  \frac{\rho_s}{\rho} = \frac{\left\langle \mathbf{W}^2 \right\rangle}{2 \lambda \beta N},
\end{equation}
where $\mathbf{W}$ is the winding number, which, in the PIMC worm framework, corresponds to the distance between the first bead of a polymer (whose position is constrained in the simulation box) and the last bead (that can occupy all space, but closes on an image of the first bead), that is
\begin{equation}\label{eq:winding}
  \mathbf{W} = \sum_{i=1}^N \sum_{j=0}^{P-1} (\mathbf{r}^{(j+1)}_i -\mathbf{r}^{(j)}_i).
\end{equation}
Notice that the winding number $\mathbf{W}$ is "quantized" in units of the box length,~\cite{pollock_1987} and the occurrence of nonzero $\mathbf{W}$ is a manifestation of superfluidity as this happens only when polymers become connected in imaginary time. 

\subsection{One-body density matrix and condensate fraction}
The one-body density matrix $\rho_1(\mathbf{r},\mathbf{r}')$
is a central quantity for BEC systems since it corresponds to the inverse Fourier transform of the momentum distribution $n(k)$.
Indeed, by definition a BEC system is determined by a macroscopic occupation of a single-particle state;~\cite{Leggett2006} in a liquid system this implies that $n(k)$ has a delta-peak for $k = 0$ and hence $ \rho_1(\mathbf{r},\mathbf{r}')$ presents a non-zero asymptotic value in the limit of large distances $|\mathbf{r} - \mathbf{r}'| \to \infty$ (the so-called off-diagonal long-range order).~\cite{penrose_1956,pitaevskii_2016} In particular, the condensate fraction $n_0$, that is the fraction of atoms inside the zero momentum state, is related to $\rho_1$ by the formula
\begin{equation}
  \lim_{|\mathbf{r}-\mathbf{r}'|\to \infty} \rho_1(\mathbf{r},\mathbf{r}') = n_0.
\end{equation}
In the context of PIMC, $\rho_1(\mathbf{q})$ with $\mathbf{q}=\mathbf{r}-\mathbf{r}'$ can be computed as
\begin{equation}
  \rho_1 (\mathbf{q}) = \frac{\Omega}{N Z} \left\langle \delta \left( \mathbf{q} - \left( \mathbf{r}^{(0)}_{i_T} - \mathbf{r}^{(P)}_{i_H} \right) \right)\right\rangle,
\end{equation}
where $\mathbf{r}^{(0)}_{i_T}$ is the position of the tail of the worm, $\mathbf{r}^{(P)}_{i_H}$ is the position of the head of the worm and the brackets here refer to the average over the configurations presenting the open polymer (G-sector). 
The condensate fraction $n_0$ is then estimated fitting the long range behaviour of $\rho_1(q)$ with the Bogoliubov expression which reads as:~\cite{boninsegni_2006_2}
\begin{equation}\label{eq:n0}
  \rho_1(q) = n_0 \ \exp\left[{\frac{T}{ \frac{4 \pi\hbar^2 \rho_s}{m} q}} \right], 
\end{equation}
where $\rho_s$ is the superfluid density in Eq.~(\ref{eq:sd}).

\section{Results}\label{sec:results}

\begin{figure}[!hbt]
  \centering
  \includegraphics[width=0.5\textwidth]{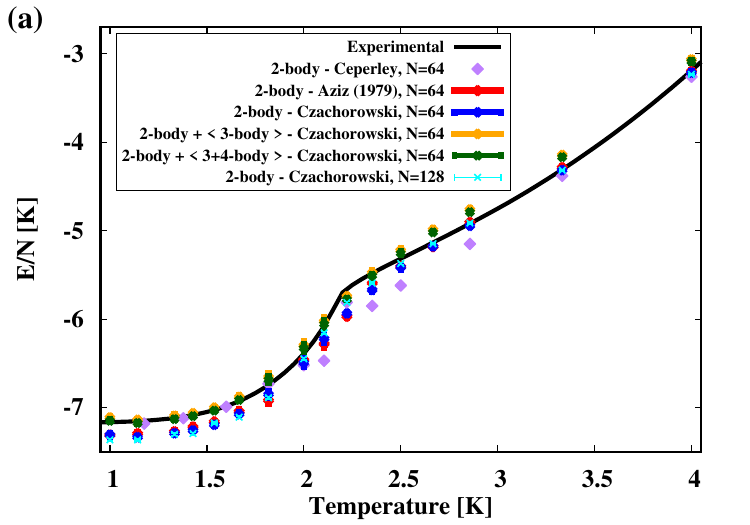}%{figures/total_energy.pdf}
  \includegraphics[width=0.49\textwidth]{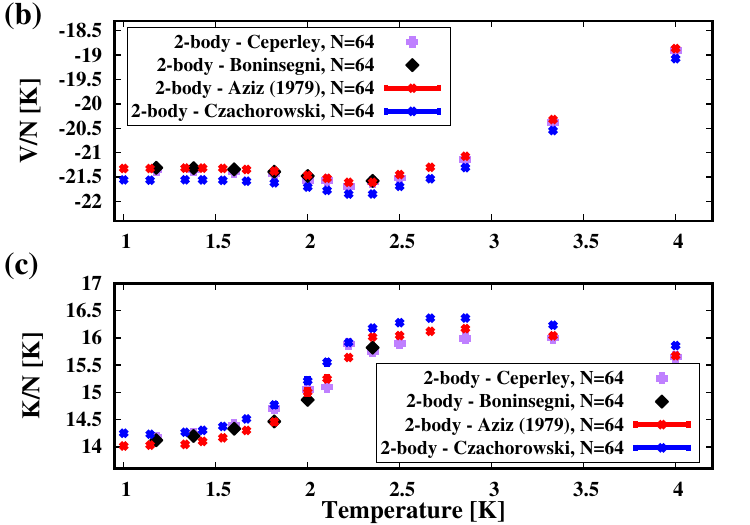}%{figures/pot_and_kin.pdf}
  \\
  \includegraphics[width=0.49\textwidth]{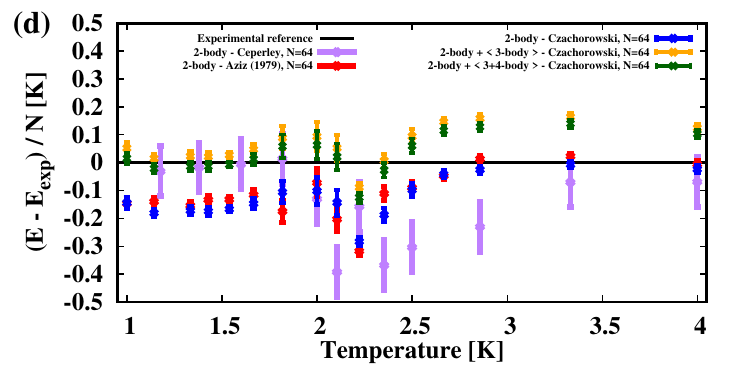}%{figures/total_energy_differences.pdf}
  \includegraphics[width=0.49\textwidth]{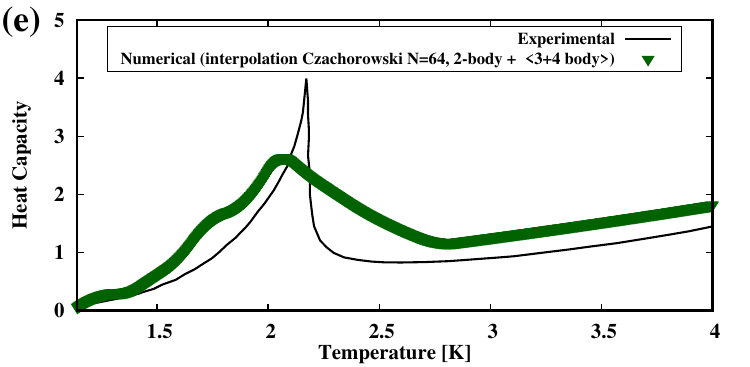}%{figures/specific_heat.pdf}
  \caption{(a) Total energy. Comparison between Aziz potential~\cite{aziz_1979} and the Czachorowski one.~\cite{czachorowski_2020} In the latter case, corrections due to 3-body (orange) and 4-body (green) have been added perturbatively. All the points have been obtained with N=64 $^4$He atoms, except for the cyan points where N=128. Experimental measurements are taken from Ref.~\cite{exp_he4} and are shifted by $-7.17$ K, which is the experimental ground state energy.~\cite{debruyn_1987} Purple points are taken from Ref.~\cite{ceperley_1986}. (b) and (c): 2-body potential and kinetic energies contributions to the total energy. Black diamonds correspond to energies from Ref.~\cite{boninsegni_2005}. (d) Differences of the N=64 results with respect to the experimental value. (e) Heat capacity. Experimental results are taken from Ref.~\cite{wilks_1967}, while the numerical points are obtained by interpolating with a cubic spline the energy plus the 3-body and 4-body contributions and differentiating the obtained function.}\label{fig:energy}
\end{figure}

The main results of our work are summarized in Fig.~\ref{fig:energy} and Tab.~\ref{tab:tab1}. Figure~\ref{fig:energy}(a) reports a comparison of our results with experimental data~\cite{exp_he4} and the original values by Ceperley and Pollock.~\cite{ceperley_1986,ceperley_1995}. 
In general, we find a good agreement, although some little differences are apparent when compared in detail with analogous calculation appeared in the literature, using various implementation of the path-integral Monte Carlo simulations and the Aziz potential from 1979.~\cite{aziz_1979}
The same panels also reports the values of the total energy computed with the pair potential by Czachorowski {\em et al.} using systems of $N=64$ and $N=128$ atoms. We do not find any difference in the results within our statistical uncertainty $u(E) \approx 0.05$~K, and therefore in the following we discuss only the results obtained with the smaller system.

Panels (b) and (c) of Fig.~\ref{fig:energy} report the average potential and kinetic energies per particle, respectively. We notice that our calculations using the Aziz potential result in average potential energies that are in very good agreement with the seminal work by Ceperley~\cite{ceperley_1986} and the original implementation of the worm algorithm~\cite{boninsegni_2006,boninsegni_2006_2} across the whole temperature range studied in this work. Additionally, our values have an overall statistical uncertainty which is half that of Ceperley, as can be seen in Tab.~\ref{tab:tab1}.
However, we find a small but systematic difference when comparing the expectation values of the
kinetic energy per particle: in this case our values are generally slightly smaller than the one
published in the literature in the superfluid phase, and slightly higher in the normal liquid phase,
with a small but systematic difference of the order of $0.1$~K. We validated our kinetic-energy
estimators by performing several simulation of the ideal gas, where the kinetic energy can be
evaluated analytically, and found an excellent agreement (see sec.~\ref{sec:freeboson}). The origin of this slight discrepancy is
not clear, but we notice that the formulation of the original worm
algorithm~\cite{boninsegni_2006,boninsegni_2006_2} was affected by an ambiguity that has been solved
in the reformulation used in the present paper.~\cite{Spada_2022} 

Also in panels (b) and (c) of Fig.~\ref{fig:energy} we report the values of the potential and kinetic energies obtained with the {\em ab initio} pair potential of Czachorowski {\em et al.}: although the total energy turns out to be the same as Aziz (within our uncertainties, see sec.~\ref{sec:azizII}) there is a difference in the breakdown of the kinetic and potential components. The potential by Czachorowski {\em et al.} results in a systematically lower potential energy and a correspondingly higher kinetic energy when compared to the Aziz potential, with a difference of $\approx 0.1$~K.

Figure~\ref{fig:energy}(d) reports the difference between our calculated energies, including the three- and four-body contributions, using the experimental values as common reference. 
In general, we notice that the effect of the three-body interaction is to shift the total energy in the positive direction by a contribution of $\approx 0.2$~K , while the contribution of the four-body potential is negative, and smaller in magnitude by almost one order of magnitude (that is, $\approx -0.02$~K). Table~\ref{tab:tab1} reports the values of the various contributions to the total energy at all the state points considered in this work.
The repulsive nature of the three-body potential in helium agrees with similar studies performed using model potentials,~\cite{boronat_1994} while to the best of our knowledge the estimate of the four-body contribution is reported here for the first time.

In the superfluid regime, our total energy values are in excellent agreement with experimental data, considering our overall statistical uncertainty of $0.05$~K. In particular, we clearly see that the inclusion of the three-body is necessary for a quantitative agreement.
However, when we consider the normal fluid, our computational results, including the three-body energy, result in a systematic overestimation of the experimental energy.
The origin of this discrepancy, in light of the very good agreement in the superfluid phase, is not
clear. Noticing that the only uncontrolled approximation used in our calculation is the pair product
form of the two-particle propagator, we tried to estimate the uncertainty introduced by this
choice. We found that the (controlled) infinite limit of the primitive approximation converges to
the actual value from below. However, simulations in the liquid phase using the primitive propagator
are very time-consuming, and the extrapolation to infinity does not allow us to confidently assess
any possible discrepancy. If such discrepancy exists, it should not exceed $\approx 0.1$~K (see
sec.~\ref{sec:pp_vs_primitive}).

\begin{table}[h!]
  \centering
  \begin{tabular}{c  c  ||  c  c  c  c  c } 
    
    $T$ [K] & \ \ density [\AA$^{-3}$] \ \ & \ \ $K/N$ [K] \ \ & $\mathcal{V}_2/N$ [K] \ \ & $ \mathcal{V}_3/N$ [K] \ \ & $ \mathcal{V}_4/N$ [K] \ \ & $(K + \mathcal{V}_2+ \mathcal{V}_3+\mathcal{V}_4)/N$ [K]   \\ [0.5ex] 
    \hline\hline
    $4.0$ & $0.01935$ & $15.858$ & $-19.073$ & $0.149$ & $-0.022$  & $-3.088$\\ 
    $3.333$ & $0.02073$ & $16.231$ & $-20.547$ & $0.171$ & $-0.024$ & $-4.169$ \\
    $2.857$ & $0.02143$ & $16.363$ & $-21.308$ & $0.184$ & $-0.030$ & $-4.791$\\
    $2.666$ &  $0.02164$ & $16.359$ & $-21.536$ & $0.188$ & $-0.031$ & $-5.020$\\
    $2.5$ & $0.02180$ & $16.277$ & $-21.692$ &  $0.193$ & $-0.034$ & $-5.256$ \\
    $2.353$ & $0.02192$ &  $16.178$ & $-21.847$ & $0.194$  &$-0.034$ & $-5.509$ \\
    $2.222$ & $0.02196$ & $15.915$ & $-21.850$ & $0.194$ & $-0.036$ & $-5.777$\\
    $2.105$ & $0.02194$ & $15.553$ & $-21.774$ & $0.196$ & $-0.033$ & $-6.057$ \\
    $2.0$ & $0.02191$ & $15.213$ & $-21.705$ & $0.196$ & $-0.031$ & $-6.327$\\
    $1.818$ & $0.02187$ & $14.771$ & $-21.620$ & $0.194$ & $-0.030$ & $-6.685$\\
    $1.666$ & $0.02185$ & $14.512$ & $-21.585$ & $0.194$ & $-0.035$ & $-6.914$\\
    $1.538$ & $0.02184$ & $14.375$ & $-21.569$ & $0.194$ & $-0.034$ & $-7.034$\\
    $1.428$ & $0.02184$ & $14.302$ & $-21.560$ & $0.195$ & $-0.034$ & $-7.097$\\
    $1.333$ & $0.02183$ & $14.268$ & $-21.557$ & $0.195$ &  $-0.036$ & $-7.123$\\
    $1.14$ & $0.02183$ & $14.232$ & $-21.566$ & $0.196$ & $-0.036$ & $-7.174$\\
    $1.0$ & $0.02183$ & $14.251$ & $-21.559$ & $0.197$ & $-0.038$ & $-7.149$\\%[1ex] 
    \hline
  \end{tabular}
  \caption{Summary of the energies using the Czachorowski potential~\cite{czachorowski_2020} and $N=64$. The standard uncertainties are $0.045$~K for the kinetic energy, $0.01$~K for the two-body potential, $0.001$~K for the three-body potential, and $2\cdot 10^{-4}$~K for the four-body potential .}
  \label{tab:tab1}
\end{table}

\begin{figure}[!hbt]
  \centering
  \includegraphics[width=0.49\textwidth]{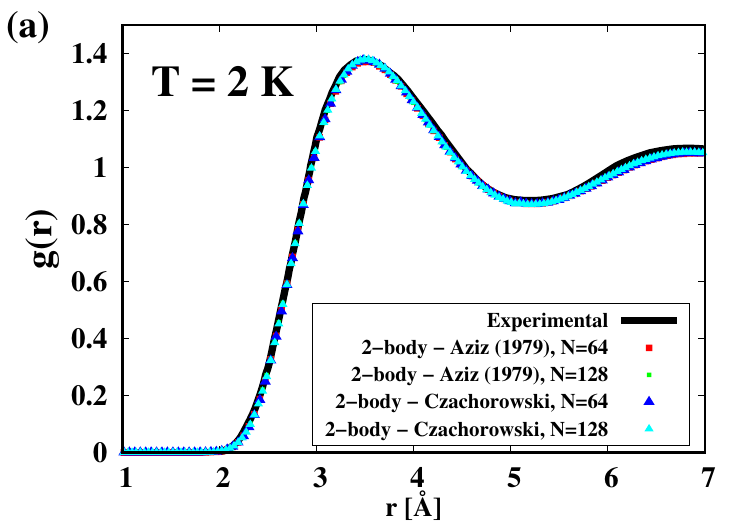}%{figures/gr_2K.pdf}
  \includegraphics[width=0.49\textwidth]{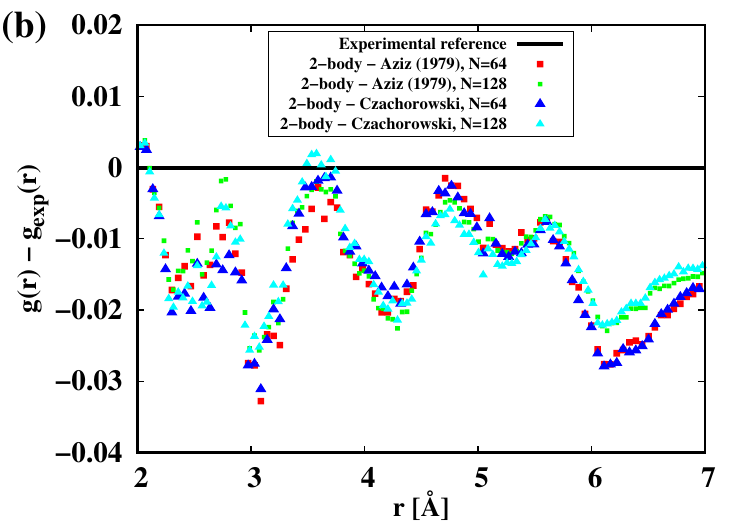}%{figures/gr_difference.pdf}
  \caption{(a) Pair distribution function at $T=2$ K for the Aziz potential~\cite{aziz_1979} and the Czachorowski one~\cite{czachorowski_2020} and for different number of particles; experimental measurements are taken from Ref.~\cite{svensson_1980}. (b) Differences of the numerical pair distribution functions with the experimental one.  \label{fig:gr_2K}}
\end{figure}

We also computed the pair correlation function, superfluid density and condensate fraction.
In Fig.~\ref{fig:gr_2K} we show the experimental measurements of the pair correlation function done at $T=2$ K using neutron-scattering techniques and the results of our simulations, using both the Aziz and Czachorowski potentials. From Fig.~\ref{fig:gr_2K}(a) we can observe that both potentials fit quite well the experimental data and that the two theoretical curves, for a given number particle, overlap almost exactly. A magnified version, showing the difference between our simulations and experimental data, is reported in Fig.~\ref{fig:gr_2K}(b). In general, all the potentials that we investigated produce compatible pair correlation functions. As expected, we find a slight difference at large values of the distance between the systems with $N=64$ and $N=128$ particles, most likely due to finite-size effects. As evidenced also in Fig.~\ref{fig:energy}(a), however, a system with $N=64$ particles already shows well-converged values of the observables.

\begin{figure}
  \centering
  \includegraphics[width=0.49\textwidth]{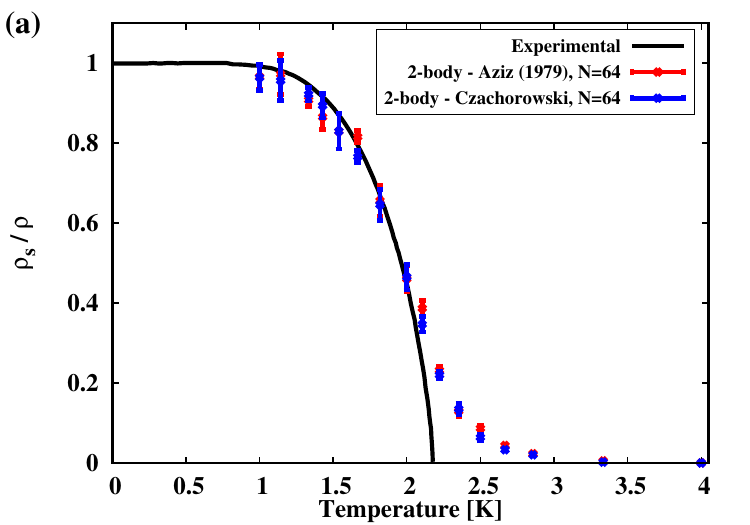}%{figures/superfluid_density.pdf}
  \includegraphics[width=0.49\textwidth]{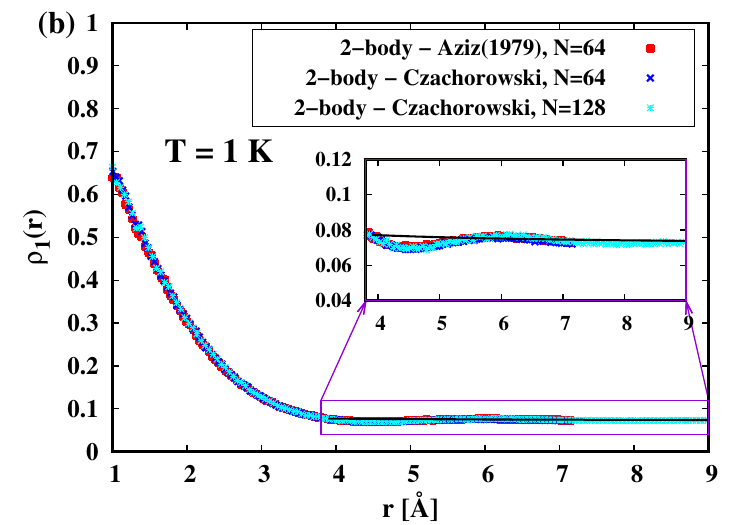}%{figures/obdm_1K.pdf}
  \caption{(a) Superfluid density. Theoretical expectation values versus experimental results. The latter are taken from Ref.~\cite{barenghi_1998}. (b) One-body density matrix. From the long-range behaviour one can extract the condensate fraction $n_0$, represented by the black line, obtained by fitting the cyan points with Eq.~(\ref{eq:n0}). The inset shows the asymptotic behaviour of $\rho_1(r)$ which coincides up to 7 \AA. The size of the tails depends on the size of the box. \label{fig:superfluid_density}}
\end{figure}

In Fig.~\ref{fig:superfluid_density}(a) we show the behaviour of the superfluid density as a function of temperature. The two pair potentials investigated here reproduce a very similar behaviour which agrees with the experimental curve except above $\sim 2$~K, where a residual superfluid density can be seen even above the superfluid transition temperature. This occurs because the finite size of the system broadens what would otherwise be a sharp transition into a smoother crossover.~\cite{pollock_1987}
Similarly, finite-size effects results in relatively large fluctuations of the superfluid density estimators below $T_{\lambda}$ and the overall uncertainty of our calculation of $\rho_s/\rho$ is $\sim 10$\% .

The one-body density matrix is shown in Fig.~\ref{fig:superfluid_density}(b). From its asymptotic behaviour one can estimate the condensate fraction $n_0$, which is equal to $n_0=0.71$ in our simulations, in good agreement with previous theoretical estimations~\cite{moroni_2004} and with experimental results.~\cite{glyde_2000}

\section{Conclusions}\label{sec:concl}

In this work, we performed extensive Path Integral Monte Carlo (PIMC) simulations of liquid helium in both the superfluid and normal phases, utilizing state-of-the-art ab initio two-body~\cite{czachorowski_2020}, three-body~\cite{lang_2023}, and four-body~\cite{wheatley_2023} interaction potentials. Our aim was to investigate how accurately the first-principles calculations, which have demonstrated remarkable precision for helium gas properties~\cite{garberoglio_2023_review}, can be extended to the liquid phase.

When compared with analogous calculations appeared in the literature, our results are in very good agreement considering our statistical uncertainty of approximately $0.05$~K. Using a perturbative approach, we demonstrated that the three-body potential contributes positively to the total energy, accounting for roughly 4\%, consistent with previous estimates based on model potentials. In contrast, the four-body potential contributes negatively, with a magnitude approximately one order smaller, representing about 0.4\% of the total energy.

When compared to experimental data, the energies per particle computed with {\em ab initio}
potentials show excellent agreement in the superfluid phase. However, we observe a systematic
discrepancy of a few percent in the normal phase. The cause of this discrepancy is unclear. In our
calculations, we employed the pair-product approximation to improve computational efficiency, which
may have introduced an uncontrolled bias. Unfortunately, comparisons between results obtained using
the pair-product approximation and the more rigorous primitive approximation were inconclusive.

Our calculations indicate that the explicit inclusion of three- and four-body potentials in the
sampling of configurations would result in significant increases in the computational cost of the
calculation. Due to the extremely slow evaluation of the four-body potential, we believe that its
explicit use in simulations is unlikely to be feasible in the foreseeable future.

Nevertheless, we believe that sampling liquid configurations using the three-body potential --
possibily with the optimized implementation discussed in Ref.~\cite{marienhagen_2024} -- and
employing higher-order non-primitive propagators~\cite{Chin23} such as those used in
Ref.~\cite{boninsegni_2005}, could be a viable approach. Apart from the advantage of incorporating
the effects of three-body forces exactly, this approach will also eliminate the uncertainties
arising from the uncontrolled approximations present in the pair-product form of the propagator.

Computer simulations with controlled approximations and explicit inclusion of the three-body
potential are expected to achieve accuracies on the order of $0.4$\%. A more accurate estimation of
the uncertainties due to the neglect of the four-body interaction will, for the time being, need to
be performed perturbatively.

\appendix

\section{Comparison between the pair-product and primitive approximations}\label{sec:pp_vs_primitive}
\begin{figure}[!h]
  \centering
  \includegraphics[width=0.89\textwidth]{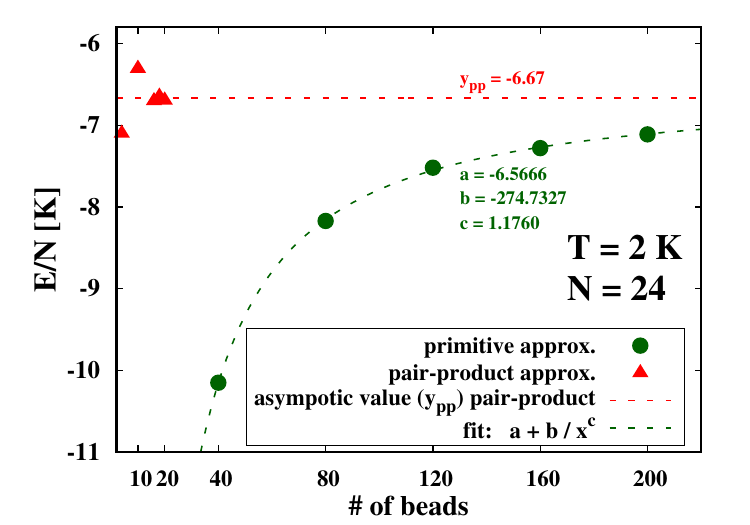}%{figures_SI/conv_beads.pdf}
  \caption{Beads convergence for the energy-per-particle for $N=24$ at $T=2$ K using the pair-product and primitive approximations.}\label{fig:convbeads}
\end{figure}

In Fig. \ref{fig:convbeads} we show the convergence of the total energy per particle for $T=2$ K and $N=24$ using the pair-product (red triangles and dashed line) and the primitive (green circles and dashed line) approximations.
We notice that while the convergence of the primitive approximation is monotonic, the results obtained from the pair-product approximation are less predictable in terms of monotonicity. This happens because the latter approach is based on an uncontrolled approximation for the propagator (Eq.~(\ref{eq:FK_approx}) of the main text). However the big advantage is that in the latter approach for $P=20$, the energy-per-particle is already converged, while using the primitive approximation convergence would be achieved around $P\sim500$.
We also notice that the energy predicted by the pair-product ($E/N=-6.67 $ K) and the primitive ($E/N=-6.567$ K) ones differ by $0.1$ K.

\begin{figure}[!h]
  \centering
  \includegraphics[width=0.89\textwidth]{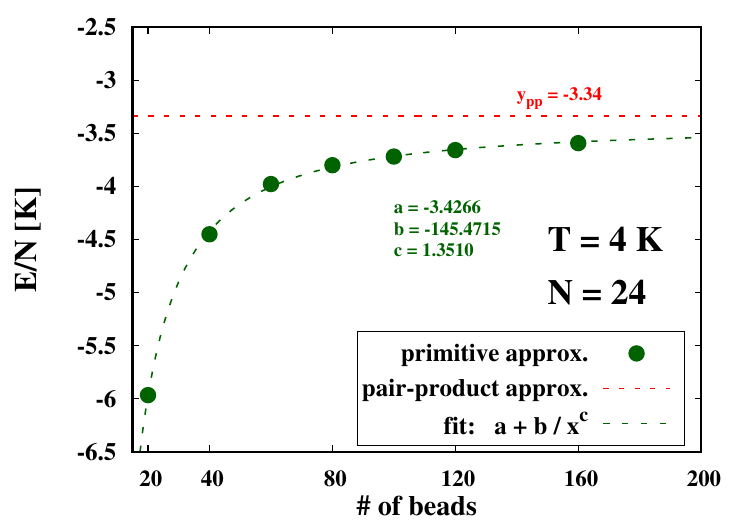}%{figures_SI/t_eq_4.pdf}
  \caption{Energy-per-particle for $N=24$ at $T=4$ K using the pair-product (red dashed line) and primitive approximations (green circles and dashed line).}\label{fig:t_eq_4}
\end{figure}

In Fig. \ref{fig:t_eq_4} we show a similar analysis but for $T=4$ K. In this plot we report directly the converged value using the pair-product (red dashed line) and we estimate the energy-per-particle of the primitive by fitting the points obtained using different number of beads. The difference between the two energies is again of the order of $0.1$ K but at variance with Fig.~\ref{fig:convbeads}, in this the primitive energy is lower than the pair-product one.

\section{Free particles: a test for the algorithm and the kinetic energy}\label{sec:freeboson}
In order to test the validity of the PIMC worm algorithm, we simulated a gas of free particles with Boson statistics and Boltzmann statistics at thermodynamical conditions similar to superfluid Helium-4. While in Ref. \cite{Spada_2022} the authors have already shown that this algorithm works perfectly in the ideal gas case, this test is useful because when the number of particles is large, kinetic energy can be troublesome to sample in the regime where quantum effects are dominating. In Tab. \ref{tab:s1} we summarize the results of this analysis. We observe that the kinetic energy of the free Helium-4 particles is always well captured independently of the thermodynamical parameters used. This means also that when the De Broglie wavelength is comparable with the size of the box, the algorithm works properly. 

\begin{table}[h]
  \centering
  \begin{tabular}{c|cc|cc}
    \toprule
    \multicolumn{1}{c}{} & \multicolumn{2}{c}{\textbf{Boltzmann} (E/N [K])} & \multicolumn{2}{c}{\textbf{Boson}  (E/N [K])} \\
    \cmidrule(rl){2-3} \cmidrule(rl){4-5}
    \textbf{Simulation parameters} & \ \ {Exact} & \ \ {Numerical} & \ \ {Exact} & \ \ {Numerical} \\
    \midrule
    N=32, P=40, $\Omega$=1466.54 \AA$^3$, T = 1 K   &  \ \ 1.3460  & \ \ 1.3451 $\pm$ 0.0015 & \ \ 0.1066  & \ \  0.1196 $\pm$ 0.0118  \\
    N=32, P=30,  $\Omega$=1466.54 \AA$^3$, T = 1.333 K & \ \ 1.9532 & \ \ 1.9536 $\pm$ 0.0010 & \ \ 0.2432 & \ \ 0.2465 $\pm$  0.0082  \\
    N=32, P=20, $\Omega$=1460.52 \AA$^3$, T = 2 K & \ \ 2.9968 & \ \ 2.9969 $\pm$ 0.0006 & \ \ 0.7194  & \ \  0.7161 $\pm$ 0.0106 \\
    N=32, P=14, $\Omega$=1493.93 \AA$^3$,T = 2.85 K & \ \ 4.2749  & \ \ 4.2750 $\pm$ 0.0010  & \ \ 1.8463 & \ \ 1.8474 $\pm$ 0.0126   \\
    N=32, P=10, $\Omega$=1656.31 \AA$^3$, T = 4 K & \ \ 6.0000 & \ \  6.0008 $\pm$  0.0010  & \ \ 4.2534 & \ \ 4.2552 $\pm$ 0.0121 \\ 
    \bottomrule
  \end{tabular}\caption{Kinetic energies per particle in the ideal gas case for different parameters considering both the Boltzmann statistics and the boson statistics.}\label{tab:s1}
\end{table}

\section{Comparison of the energies with another different semi-empirical potential}\label{sec:azizII}
\begin{figure}[!h]
  \centering
  \includegraphics[width=0.89\textwidth]{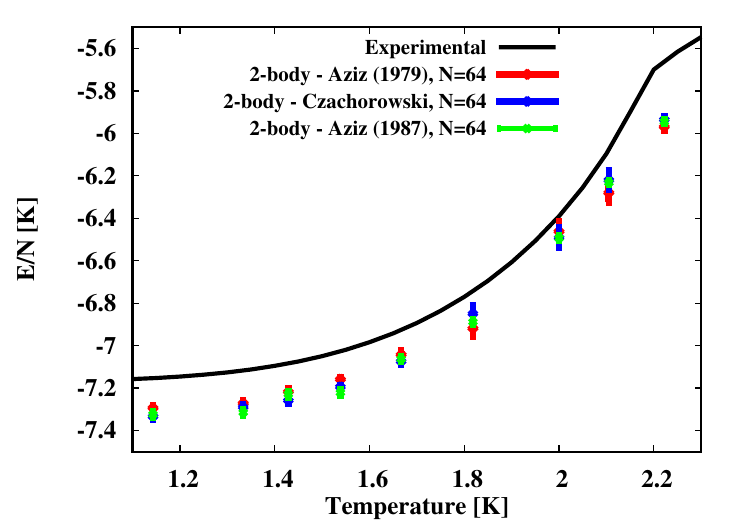}%{figures_SI/total_energy_azizII.pdf}
  \caption{Total energy per particle at low T computed with the same potentials of the main text plus another semiempirical potential \cite{aziz_1987}}\label{fig:azizII}
\end{figure}

In Fig.~\ref{fig:azizII} we show the energy-per-particle as in Fig.~\ref{fig:energy}(a) of the main
text, using only pair-potentials (i.e. without corrections of 3- and 4-body potentials). In
particular one can see that using another semi-empirical potential \cite{aziz_1987} we get the same
total energy as using the two potentials reported in the manuscript. 

For the sake of comparison and reproducibility, in Tab.~\ref{tab:tab_aziz79}, we also summarize the energies using the Aziz-1979 potential.

\begin{table}[h!]
  \centering
  \begin{tabular}{c  c  ||  c  c  c  c } 
    $T$ [K] & \ \ density [\AA$^{-3}$] \ \ & \ \ $K/N$ [K] \ \ & $\mathcal{V}_2/N$ [K] \ \ & $(K + \mathcal{V}_2)/N$ [K]   \\ [0.5ex] 
    \hline\hline
    $4.0$ & $0.01935$ & $15.671$ & $-18.868$ &  $-3.196$\\ 
    $3.333$ & $0.02073$ & $16.037$ & $-20.323$ &  $-4.286$ \\
    $2.857$ & $0.02143$ & $16.169$ & $-21.075$ &  $-4.906$\\
    $2.666$ &  $0.02164$ & $16.119$ & $-21.301$ &  $-5.182$\\
    $2.5$ & $0.02180$ & $16.044$ & $-21.450$ &   $-5.406$ \\
    $2.353$ & $0.02192$ &  $16.016$ & $-21.609$ &  $-5.592$ \\
    $2.222$ & $0.02196$ & $15.640$ & $-21.610$ &  $-5.970$\\
    $2.105$ & $0.02194$ & $15.248$ & $-21.528$ &  $-6.279$ \\
    $2.0$ & $0.02191$ & $15.003$ & $-21.466$ &  $-6.462$\\
    $1.818$ & $0.02187$ & $14.460$ & $-21.379$ &  $-6.918$\\
    $1.666$ & $0.02185$ & $14.303$ & $-21.346$ &  $-7.042$\\
    $1.538$ & $0.02184$ & $14.167$ & $-21.325$ &  $-7.159$\\
    $1.428$ & $0.02184$ & $14.103$ & $-21.320$ &  $ -7.216$\\
    $1.333$ & $0.02183$ & $14.047$ & $-21.318$ &  $-7.270$\\
    $1.14$ & $0.02183$ & $14.034$ & $-21.327$ &  $ -7.293$\\
    $1.0$ & $0.02183$ & $14.017$ & $-21.327$ &  $-7.310$\\%[1ex] 
    \hline
  \end{tabular}
  \caption{Summary of the energies using the Aziz potential~\cite{aziz_1979} and $N=64$. The standard uncertainties are $0.048$~K for the kinetic energy and $0.01$~K for the two-body potential.}
  \label{tab:tab_aziz79}
\end{table}

Finally, we also report the difference of the total-energy-per-particle with the experimental curve, as in Fig.~\ref{fig:energy}(d) of the manuscript, where we collect all the results with 2-body potentials. 
\begin{figure}
  \centering
  \includegraphics[width=0.99\linewidth]{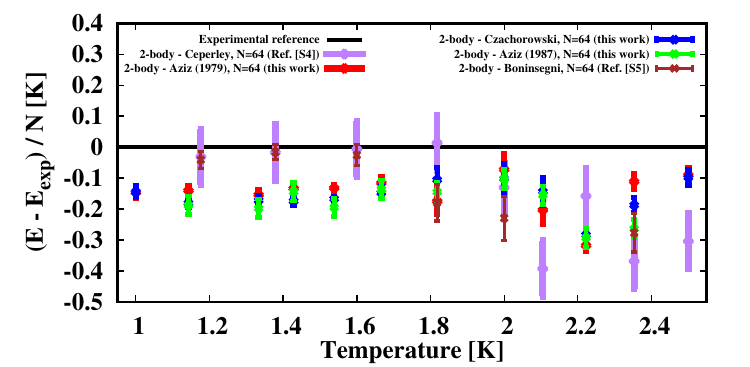}%{figures_SI/total_energy_differences_supp.pdf}
  \caption{Differences of the N=64 results with respect to the experimental value (black line). Ceperley results are taken from Ref.~\cite{ceperley_1986} while Boninsegni results from Ref.~\cite{boninsegni_2005}.}
  \label{fig:ediff_supp}
\end{figure}

\clearpage
\section*{Acknowledgements}
TM and GG thank Gabriele Spada and Stefano Giorgini of the Department of Physics of the University
of Trento for useful discussions, expecially on the details of the PIMC+worm algorithm.
\section*{Author Contributions} 
All authors developed the theory and analysed the data.
TM coded the numerical algorithms and performed the calculations. 
TM and GG wrote the paper.
\section*{Competing Interests}
The authors declare no competing interests.
\section*{Data Availability}
The data that support the findings of this study are available from the corresponding author upon reasonable request.
\section*{Code Availability}
The codes implementing the calculations of this study are available from the corresponding author upon request.

\clearpage

%\bibliography{biblio}

%% BioMed_Central_Bib_Style_v1.01

\end{document}